# Forecasting U.S. Textile Comparative Advantage Using Autoregressive Integrated Moving Average Models and Time Series Outlier Analysis

Zahra Saki[1], Lori Rothenberg[1], Marguerite Moore[1]
Ivan Kandilov[2], Blanton Godfrey[1]
[1]Department of Textile Technology Management, College of Textiles,
[2]Department of Agricultural and Resource Economics,
NC State University, Raleigh, NC

**Abstract**
To establish an updated understanding of the U.S. textile and apparel (TAP) industry's competitive position within the global textile environment, trade data from UN-COMTRADE (1996-2016) was used to calculate the *Normalized Revealed Comparative Advantage* (NRCA) index for 169 TAP categories at the four-digit Harmonized Schedule (HS) code level. Univariate time series using *Autoregressive Integrated Moving Average* (ARIMA) models forecast short-term future performance of *Revealed* categories with export advantage. Accompanying outlier analysis examined permanent level shifts that might convey important information about policy changes, influential drivers and random events.

## 1. Introduction

Forecasting represents an important area of enquiry in business and academic research. Practitioners in sales, marketing, supply chain and other fields advocate the importance of forecasting for business success (Dalrymple, 1987; Fildes and Hastings, 1994; Bendato et al., 2015). Additionally, economic forecasting has always been an integral part of policymaking for individual countries (Sims, 1986).

In 1786 Playfair first illustrated the historical economic performance of England using import and export data (Tufte, 2001, p. 32). Much later, the field of economics introduced the theory of comparative advantage to predict trade patterns among countries. In 1965 Balassa developed a new approach to compare and predict trade patterns from a global perspective. His index referred to as *Revealed Comparative Advantage* (RCA) provides an indicator of competitiveness based on historical trade data. Since its introduction, researchers modified RCA (Vollrath, 1991; Hoen and Oosterhaven, 2006) to overcome various shortcomings of the index. One of the most recent modifications, known as *Normalised Revealed Comparative Advantage* (NRCA) (Yu et al., 2009) claims stable distribution over time and enables application of time-series analysis to comparative advantage.

Recent policies and initiatives encourage U.S. TAP domestic manufacturing. This trend emphasizes the need to understand the past in order to forecast the future. While most research focuses, on TAP products that lost comparative advantage due to globalization, this enquiry focuses on U.S. TAP products that maintained export comparative advantage in recent years (e.g., at least three consecutive years between 2010-2016).

Existing academic research into U.S. TAP competitiveness that uses RCA or its variants rely on factor endowment analysis and graphical illustration of indices values over time (Chi et al., 2005; Chi and Kilduff, 2006; Kilduff and Chi, 2006b; Kilduff and Chi, 2006a). Outdated research and the past tendency to rely on qualitative approaches to times series analysis suggest the necessity for an updated investigation into competitiveness of U.S. TAP products using quantitative time series analysis. Therefore, the purpose of this study is to utilize univariate time series analysis based on *Autoregressive Integrated Moving Average* (ARIMA) to forecast short-term performance of *Revealed* U.S. TAP categories with export advantage (i.e., six categories). The corresponding ARIMA outlier analysis correlates past shifts in trends to identify possible drivers and random events. The research examines *Normalized Revealed Comparative Advantage* (NRCA) for 169 TAP categories at the four-digit HS code level (1996-2016). The following objectives address the research purpose:

> Research Objective 1 (RO1): To determine NRCA (four-digit harmonized code) among U.S. TAP categories from 1996-2016 (21 years) and identify categories that indicate comparative advantage over a minimum of three consecutive years between 2010 and 2016.
>
> Research Objective 2 (RO2): To identify appropriate time series models to forecast short-term (2-years) comparative advantage among categories identified in RO1.
>
> Research Objective 3 (RO3): To identify significant changes in trends among categories by interpreting additive outliers and permanent level shifts generated by the time-series models.

Forecasting the competitive position of TAP categories identified as competitive in recent years (at least three consecutive years between 2010 and 2016) is necessary for practitioners to assess opportunities and invest resources. The categories that forecast competitive decline can signal calls to action for practitioners and policymakers alike. Additionally, insight from outlier analysis identifies potentially important level shifts in TAP categories' competitiveness.

## 2. Data Description

The United Nations Commodity Trade database (UN COMTRADE) provides data for the study. The database contains annual bilateral merchandise trade metrics (i.e., imports and exports) at the country level for different versions of standardized coding systems. The metrics for Harmonized Commodity Description and Coding System (HS) use two, four and six digit product classifications, which provide increasingly specific product information, respectively. Product-level trade data on the 1996 revision at four-digit HS code levels from 1996 to 2016 constitutes the data for this study[1]. The dataset includes Chapters 50-67 (i.e., textiles, textile articles, footwear, etc.).

After compiling the data, NRCA is calculated to identify products with comparative advantage. NRCA values for *Revealed* categories from 1996 to 2015 comprise the ARIMA training models and NRCA values for 2016 test these models.

---

[1] World Integrated Trade Solution (2016)

## 3. Measures and Methodologies

### 3.1 NRCA

To predict the trade pattern among countries, the classical theory of comparative advantage assumes production cost as the sole indicator of a given country's specialization, which does not account for non-price factors (e.g., consumer preference, product quality) that impact trade flows. To address these limitations, Balassa introduced the concept of *Revealed Comparative Advantage* (RCA) (1965) which predicts trade patterns using historical trade data. RCA values vary from zero to infinity, which makes the index asymmetric. Therefore, empirical and theoretical research identifies inherent shortcomings of the index (i.e., Proudman and Redding, 2000; De Benedictis and Tamberi, 2001; Hoen and Oosterhaven, 2006). Yu et al. (2009) derived a recent variant of RCA known as *Normalized Revealed Comparative Advantage* (NRCA). The NRCA is defined as:

$$NRCA = (E_j^i/E) - (E_j * E^i/E * E) \qquad (1)$$

Where $E_j^i$ stands for the country i export of commodity j,
$E_j$ refers to the export of commodity j by all countries in the world,
$E^i$ is the country i export of all commodities,
and $E$ is the export of all commodities by all countries.
The neutral point of NRCA is zero, therefore deviations from zero indicate a country's comparative advantage or disadvantage in a given commodity. An additional feature of NRCA is its ability to generate stable distributions over time, which allows application of time-series analysis to comparative advantage.

The NRCA at the four-digit level generates the metrics to identify specific textile categories (i.e., 169 categories) with export comparative advantage over the full 21-year period. Further analysis of the indices in a focused timeframe identifies products that indicate a minimum of three consecutive years of comparative advantage in the recent past (2010-2016) (RO1).

### 3.2 ARIMA

Box and Jenkins (1976) developed a method, ARIMA, for analysing stationary time series data. The ARIMA method differences the series to stationary and combines autoregressive parameters to moving average. An autoregressive parameter, AR, indicates that the value of the series during the current period is a function of its immediate previous values and some error, while the moving average parameter, MA, involves a finite memory of past time lags. The order of MA indicates the number of time lags. The ARIMA model is capable of greater flexibility and power compared to both extrapolative and decomposition models. The Box and Jenkins model requires discrete, equally spaced data with no missing values. The series should be or made to be stationary for a time-invariant model. A stationary series presents stable but rapidly decreasing autocorrelation whereas a non-stationary series diminishes auto-correlation gradually.

Prior to fitting the time series models, the Augmented Dickey Fuller Unit Root (ADF) test is used to identify differencing orders and build a stationary series. Using the stationary series three different methods, Extended Sample Autocorrelation Function (ESACF), Minimum Information Criterion (MINIC), and The Smallest Canonical (SCAN), which tentatively identify the order of ARMA process, suggest different options for the lag order of AR and/or MA terms, p and q respectively. Suggested p and q orders generate several

ARIMA models. Akaike's Information Criterion (AIC) evaluates model fit and identifies the best p and q order. Additionally, the χ2 statistic examines residuals to assure adequate model fit. Finally, fitted models generate two-year forecasts for NRCA (2017 and 2018).

**3.3 Additive Outlier and Permanent Level Shift**
Outlier analysis is typically used to detect and remove anomalous observations. Researchers address the importance of outliers and debate whether they should be kept or removed (Osborne and Overbay, 2004). For the purpose of this study, outlier analysis identifies possible indicators of losing or gaining comparative advantage. Therefore, outlier analysis is performed along with time series analysis to detect the shifts in level or additive outliers to the response series that are not accounted for by the previously estimated model. Specifically the model considers permanent level shifts because they can convey important information about policy changes or other influential drivers and random events.

## 4. Results and Discussion

NRCA at the four-digit level identifies specific textile categories (i.e., 169 categories) with export comparative advantage over the full 21-year period. Interpretation of the indices suggests six categories that meet the requirement of three sustained years of appreciable comparative advantage. The categories that indicate adequate advantage include; *HS5201* (Cotton; not carded or combed), *HS5502* (Artificial filament tow), *HS5603* (Nonwovens; whether or not impregnated, coated, covered or laminated), *HS5205* (Cotton yarn (other than sewing thread), containing 85% or more by weight of cotton, not put up for retail sale), *HS5703* (Carpets and other textile floor coverings; tufted, whether or not made up), and *HS6309* (Textiles; worn clothing and other worn articles). See Table 1 for NRCA values of revealed categories 1996-2016 (RO1).

Table two presents the Augmented Dickey Fuller Unit Root (ADF) tests for revealed products from which cotton fiber (HS5201), nonwovens (HS5603), cotton yarn (HS5205) and worn clothing (HS6309) requires differentiation to make stationary series. The ADF test is repeated to ensure that first order differentiation effectively made the series stationary and no additional differentiation is needed (Table 3).

Table four presents the tentative models with the order of p + d and q using the SCAN and ESACF methods and p, q order using the MINIC method. The suggested orders of p and q are applied to the training dataset. At this point the model indicating the minimum AIC, is selected to fit the data(RO2) (Table 5). Based on the minimum AIC criterion in four out of six categories one order of differentiation was the only required element in the time series (I(1)). That is, the suggested order of p and q relevant to autoregressive and moving average is equal to zero. I(1) series represents white noise after differencing and is formulated as $NRCA_t = NRCA_{t-1} + e_t$. Artificial filament tow, HS5502, involves an autoregressive parameter with the order of two and carpet and other floor covering, HS5703, requires one autoregressive parameter.

Table 1: NRCA for TAP at Four-Digit Level, Indicated a Minimum of Three Consecutive Years of Comparative Advantage (2010-2016)

| Year | Cotton fiber 5201 | Artificial filament tow 5502 | Nonwovens 5603 | Cotton yarn >85% 5205 | Carpet 5703 | Worn Clothing 6309 |
|---|---|---|---|---|---|---|
| 1996 | 474.32 | 73.14 | 3.91 | -102.75 | -8.37 | 23.11 |
| 1997 | 420.97 | 53.29 | 18.69 | -90.44 | 7.80 | 22.84 |
| 1998 | 392.00 | 53.07 | 15.41 | -75.10 | 11.50 | 17.54 |
| 1999 | 92.27 | 47.23 | 9.92 | -61.45 | 11.22 | 13.01 |
| 2000 | 217.65 | 35.92 | 26.78 | -44.77 | 22.18 | 15.24 |
| 2001 | 273.19 | 45.73 | 32.63 | -42.71 | 21.61 | 12.79 |
| 2002 | 255.00 | 40.32 | 39.27 | -32.39 | 19.10 | 13.69 |
| 2003 | 366.25 | 34.13 | 50.53 | -32.45 | 18.00 | 15.00 |
| 2004 | 406.10 | 37.40 | 50.64 | -14.37 | 22.47 | 15.49 |
| 2005 | 334.84 | 39.64 | 61.20 | 1.69 | 25.70 | 13.20 |
| 2006 | 332.55 | 37.25 | 60.09 | 5.98 | 26.17 | 10.89 |
| 2007 | 293.19 | 40.76 | 39.51 | 11.12 | 17.13 | 12.44 |
| 2008 | 269.48 | 45.10 | 39.46 | 17.85 | 24.97 | 13.81 |
| 2009 | 237.20 | 61.84 | 41.84 | 20.24 | 22.68 | 14.16 |
| 2010 | 326.33 | 50.86 | 45.75 | 15.74 | 24.73 | 15.51 |
| 2011 | 400.89 | 44.47 | 39.32 | 50.69 | 22.07 | 18.34 |
| 2012 | 277.48 | 50.92 | 42.85 | 18.85 | 23.23 | 18.59 |
| 2013 | 241.65 | 52.98 | 44.10 | 7.87 | 20.43 | 19.26 |
| 2014 | 189.75 | 51.94 | 40.45 | 12.02 | 19.05 | 18.47 |
| 2015 | 198.94 | 47.55 | 36.44 | 13.06 | 16.50 | 16.17 |
| 2016 | 239.33 | 50.07 | 33.56 | 20.09 | 17.20 | 17.82 |

Extreme advantage 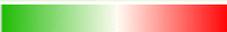 Extreme Disadvantage

*Note: NRCA values are multiplied by 10^6

Table 2: Augmented Dickey Fuller Unit Root (ADF) single mean test results (The null hypothesis is the non-stationary series ($\alpha \leq 0.05$).

| Categories | Tau | Pr < Tau | Stationary |
|---|---|---|---|
| 5201 | -2.93 | 0.0600 | No |
| 5502 | -3.75 | 0.0119 | Yes |
| 5603 | -2.44 | 0.1456 | No |
| 5205 | -2.35 | 0.1667 | No |
| 5703 | -5.29 | 0.0005 | Yes |
| 6309 | -2.53 | 0.1247 | No |

Table 3: Augmented Dickey Fuller Unit Root (ADF) single mean test results after first order differentiation (The null hypothesis is the non-stationary series ($\alpha \leq 0.05$).

| Categories | Tau | Pr < Tau | Stationary | Period of Differencing |
|---|---|---|---|---|
| HS5502 | -4.63 | 0.0021 | Yes | 1 |
| HS5603 | -4.10 | 0.0061 | Yes | 1 |
| HS5205 | -4.70 | 0.0018 | Yes | 1 |
| HS6309 | -3.19 | 0.0375 | Yes | 1 |

**Table 4:** Tentative model order selection using SCAN, ESACF and MINIC

|            | SCAN |   |      | ESACF |   |        | MINIC |   |        |
|------------|------|---|------|-------|---|--------|-------|---|--------|
| Categories | p+d  | q | BIC  | p+d   | q | BIC    | p     | q | BIC    |
| HS5201     | 0    | 0 | 7.06 | 0     | 0 | 7.06   | 0     | 3 | -30.36 |
|            | -    | - | -    | 1     | 0 | 7.20   | -     | - | -      |
| HS5502     | 0    | 1 | 3.02 | 0     | 0 | 3.03   | 3     | 1 | -36.49 |
|            | -    | - | -    | 2     | 0 | 1.85   | -     | - | -      |
|            | -    | - | -    | 3     | 0 | -31.36 | -     | - | -      |
| HS5603     | 0    | 0 | 3.57 | 0     | 0 | 3.57   | 1     | 3 | -35.79 |
| HS5205     | 0    | 0 | 2.77 | 0     | 0 | 2.77   | 4     | 2 | -36.06 |
|            | -    | - | -    | 1     | 0 | 2.86   | -     | - | -      |
|            | -    | - | -    | 2     | 0 | 2.84   | -     | - | -      |
| HS5703     | 1    | 0 | 0.78 | 1     | 0 | 0.78   | 1     | 2 | -33.87 |
|            | 0    | 1 | 1.06 | 0     | 1 | 1.06   | -     | - | -      |
| HS6309     | 0    | 0 | -0.24| 0     | 0 | -0.24  | 3     | 2 | -37.99 |
|            | -    | - | -    | 1     | 0 | -0.15  | -     | - | -      |
|            | -    | - | -    | 2     | 0 | -2.68  | -     | - | -      |

**Table 5:** Choosing the order of p, d, and q with minimum AIC
(* did not converge)

| Categories | p | d | q | AIC    |
|------------|---|---|---|--------|
| HS5201     | 0 | 1 | 0 | 228.08 |
| HS5502     | 2 | 0 | 0 | 145.50 |
| HS5603     | 0 | 1 | 0 | 136.72 |
| HS5205     | 0 | 1 | 0 | 154.09 |
| HS5703*    | 1 | 0 | 0 | 129.14 |
| HS6309     | 0 | 1 | 0 | 85.71  |

The χ2 test statistics fail to reject the no-autocorrelation hypothesis at an alpha of 0.05 indicating that the residuals are white noise, therefore the applied models are adequate for all series (Table 6). To evaluate model accuracy, forecast error is calculated by comparing actual and forecasted NRCA values for 2016 (Table 7). The largest forecast error equal to 22.93 percent belongs to cotton fiber followed by nonwovens and worn clothing. The forecast error for artificial filament tow, cotton yarn and carpet is less than five percent. Assuming observed errors as reasonable, NRCA values are forecasted for 2017 and 2018 (see Table 8). Overall, the two-year forecast for cotton fiber, and worn clothing (HS5201, HS5603) decreases compared to the actual NRCA value of 2016. Nonwovens and cotton yarn (HS5603, HS5205) shows an increase in NRCA value. The NRCA changes for artificial filament tow (HS5502) are not considerable, however this value decreases for 2017 and increases for 2018. Carpet (HS5703) shows a slight decrease for 2017 and remains the same for 2018 compared to 2017.

**Table 6:** Autocorrelation check for residuals Significance level α=0.05
(Hypothesis: Residuals are white noise)

| Categories | ARIMA model | To lag | Chi-Square | DF | Pr > ChiSq |
|------------|-------------|--------|------------|----|-----------| 
| HS5201     | (0,1,0)     | 6      | 3.07       | 6  | 0.7997    |
| HS5502     | (2,0,0)     | 6      | 2.34       | 4  | 0.6731    |
| HS5603 | (0,1,0) | 6 | 2.36 | 6 | 0.8836 |
| HS5205 | (0,1,0) | 6 | 1.61 | 6 | 0.9519 |
| HS5703 | (1,0,0) | 6 | 1.69 | 5 | 0.8896 |
| HS6309 | (0,1,0) | 6 | 2.55 | 6 | 0.8632 |

**Table 7:** NRCA Forecast for 2016 and Forecast Error

| Categories | Actual 2016 | Forecast 2016 | Std. Error | 95% Confidence Limits | | Percent Forecast Error |
|---|---|---|---|---|---|---|
| HS5201 | 239.327 | 184.445 | 95.356 | -2.449 | 371.340 | 22.93 |
| HS5502 | 50.073 | 48.808 | 8.583 | 31.986 | 65.631 | 2.53 |
| HS5603 | 33.561 | 38.153 | 8.613 | 21.273 | 55.033 | 13.68 |
| HS5205 | 20.087 | 19.151 | 13.604 | -7.513 | 45.815 | 4.66 |
| HS5703 | 17.195 | 16.504 | 5.826 | 5.085 | 27.922 | 4.02 |
| HS6309 | 17.815 | 15.802 | 2.250 | 11.391 | 20.212 | 11.30 |

**Table 8**: NRCA forecast for 2017 and 2018

| HS Code | 2017 Forecast | | | | 2018 Forecast | | | |
|---|---|---|---|---|---|---|---|---|
| | Forecast | Std Error | 95% Confidence Limits | | Forecast | Std Error | 95% Confidence Limits | |
| 5201 | 169.95 ▼ | 134.85 | -94.36 | 434.26 | 155.46 ▼ | 165.16 | -168.25 | 479.16 |
| 5502 | 49.17 ▼ | 11.10 | 27.41 | 70.92 | 49.66 ▲ | 13.17 | 23.85 | 75.46 |
| 5603 | 39.87 ▲ | 12.18 | 15.99 | 63.74 | 41.58 ▲ | 14.92 | 12.34 | 70.81 |
| 5205 | 25.25 ▲ | 19.24 | -12.46 | 62.95 | 31.34 ▲ | 23.56 | -14.84 | 77.52 |
| 5703 | 16.50 ▼ | 8.24 | 0.36 | 32.65 | 16.50 ■ | 10.09 | -3.27 | 36.28 |
| 6309 | 15.44 ▼ | 3.18 | 9.20 | 21.67 | 15.07 ▼ | 3.90 | 7.43 | 22.71 |

The additive outlier and permanent level shift analysis performed along with the time series indicates one outlier for all categories with the exception of worn clothing (HS6309) (Table 7). Both permanent level shifts and additive outliers are illustrated on forecast graphs (Figure 1). Permanent level shifts for artificial filament tow and carpet after 1997 might be associated with the WTO phase I quota restriction elimination or implementation of the North American Free Trade Agreement (NAFTA) in 1994. Specifically for carpet, which is mostly traded regionally due to its bulkiness (extensive shipping cost per square meter) the implementation of NAFTA in 1994 may have caused an increase in favor of U.S. carpet export competitiveness. A permanent level shift for nonwovens occurred in 2007 which, clearly shows a stop point to its consistent growth from 1998. Nonwovens NRCA dropped by more than 30 percent in 2007. Further investigation into gross U.S. nonwovens exports shows that the category is not actually declining in terms of export value. This suggests that emerging technologies in other countries (increased nonwoven export competition) contribute to the observed NRCA decrease in 2007. Specifically, increasing gross exports of nonwovens from China explains this observation. Chinese exports of HS5603 increased 30 percent in 2007 and quintupled in 2016 compared to 2006. Outlier analysis of cotton fiber (HS5201) only suggests one additive outlier in 1999. Additive outliers are associated with random events and do not relate to structural changes in series. The abrupt drop of cotton fiber competitiveness in 1999 is commonly explained by the 1998 drought (Outlook for U.S. Agricultural Exports, 1998) which led to less yield and exports. An additional additive outlier is identified in 2011 for cotton yarn. Further investigation into this finding does not yield a viable explanation.

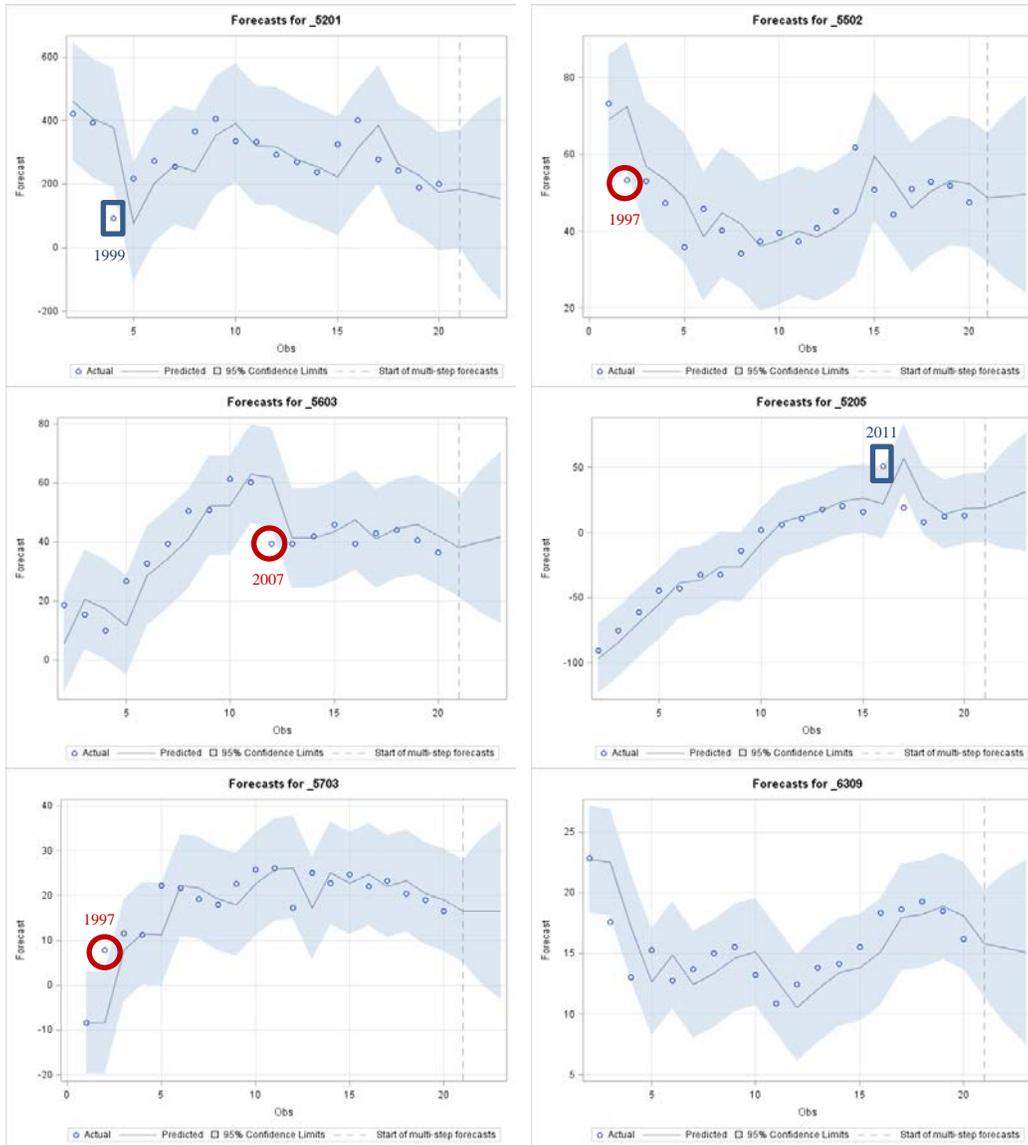

**Figure 1:** Forecast graph indicating outliers for *Revealed* categories
(Observations indicated by circles are level shifts and by rectangles are additive outliers)

**Table 9:** Additive Outlier and Permanent Level Shifts

| Categories | Additive Outlier | Permanent Level Shift |
|---|---|---|
| HS5201 | 1999 | - |
| HS5502 | - | 1997 |
| HS5603 | - | 2007 |
| HS5205 | 2011 | |
| HS5703 | - | 1997 |
| HS6309 | - | - |

## 5. Conclusion, Future Research and Limitations

This research demonstrates a first time application of the ARIMA procedure to forecast U.S. TAP export competitiveness using NRCA. NRCA at a four-digit level reveals six categories (i.e., cotton fiber, artificial filament tow, nonwovens, cotton yarn, carpet, and worn clothing) with recent sustained export advantage that creates the basis for further time series analysis to forecast short-term future and identify outliers. Cotton fiber, the most important source of U.S. TAP advantage, is forecasted to lose advantage in 2017 and 2018 compared to 2016. However, high forecast error (i.e., more than 22 percent) suggests examination of additional times-series method such as cyclical approaches. Export advantage projection of artificial filament tow which is driven by the availability of resources (mainly cellulose from wood) suggests a slight decline in 2017 and an increase in 2018. However, the magnitude of the change is negligible.

An increasing trend of nonwovens export advantage stopped in 2007 after which NRCA value oscillates at a reduced NRCA value. This observation signals increasing competition from emerging economies in technical and knowledge intensive products. Cotton yarn is projected to continue to increase advantage for 2017 and 2018. Export advantage of carpet and other floor covering is expected to experience a slight decline in 2017 and maintain a similar level in 2018. Export advantage of worn clothing, more of a challenge than a source of advantage, is expected to decline in the next two years. Further, application of outlier analysis to identify permanent level shifts and additive outliers, and correlate those to influential drivers and random events, provides an innovative method that not only improves the accuracy of models but also conveys valuable information about the sources of loosing or gaining export advantage. The most important insight from the outlier analysis is the permanent level shift of Nonwovens in 2007. Although, this observation does not necessarily relate to changes in policy and suggest further investigation. Due to the relative nature of NRCA (if a country loses advantage other countries gain advantage) and further analysis of trade data for the nonwoven industry indicate that Chinese export growth of nonwovens contributes to U.S. declining advantage.

A limitation of this research is the sole analysis of products with comparative advantage. In addition to this enquiry, further analysis of U.S. TAP products that lost advantage using the NRCA approach is insightful. The results suggest high forecast error percent for the cotton fiber, nonwovens and worn clothing which requires further investigation and application of other time series methods (i.e, cyclical). Another limitation relevant to *revealed comparative* advantage is its sole reliance on exports as an indicator of competitiveness.

In terms of measurement, the small number of data points, 21, can affect the accuracy and reliability of the ARIMA process. On the other hand, compiling NRCA using data before 1996 creates other sources of error such as the existence of different HS code versions (revisions) which necessitated merging of new and old categories over time. Second, trade date prior to 1996 is not inclusive of all countries' trade activity which can result in less reliable NRCA values.

JSM 2018 - Business and Economic Statistics Section